\documentclass[12pt]{iopart}
\usepackage[english]{babel}
\usepackage{graphicx}
\usepackage{color}
\usepackage{hyperref}
    \hypersetup{colorlinks,linkcolor=blue,citecolor=blue,urlcolor=blue}

\begin{document}

\title{Bragg spectroscopy and Ramsey interferometry with an ultracold Fermi gas}

\author{B Deh, C Marzok, S Slama, C Zimmermann, and Ph W Courteille}

\address{Physikalisches Institut, Eberhard-Karls-Universit\"at T\"ubingen,
    \\Auf der Morgenstelle 14, D-72076 T\"ubingen, Germany}

\ead{deh@pit.physik.uni-tuebingen.de}

\begin{abstract}
We report on the observation of Bragg scattering of an ultracold Fermi gas of $^6$Li atoms at a
dynamic optical potential. The momentum states produced in this way oscillate in the trap for time
scales on the order of seconds, nearly unperturbed by collisions, which are absent for ultracold
fermions due to the Pauli principle. In contrast, interactions in a mixture with $^{87}$Rb atoms
lead to rapid damping. The coherence of these states is demonstrated by Ramsey-type matter wave
interferometry. The signal is improved using an echo pulse sequence, allowing us to observe
coherence times longer than 100\,$\mu$s. Finally we use Bragg spectroscopy to measure the {\em
in-situ} momentum distribution of the $^6$Li cloud. Signatures for the degeneracy of the Fermi gas
can be observed directly from the momentum distribution of the atoms inside the trap.
\end{abstract}


\tableofcontents

\section{Introduction}

Bragg diffraction of cold atoms at propagating standing light waves is used in two distinct ways, as
Bragg spectroscopy to measure the momentum distribution of ultracold gases \cite{Stenger99} and as
beamsplitter for coherent atom optics \cite{Kozuma99}. Bragg spectroscopy has developed into a
powerful tool for measuring the dispersion relation and the dynamic structure factor of cold gases.
It was used to observe the momentum distribution of trapped Bose-Einstein condensates (BEC)
\cite{Stenger99,Stamper-Kurn99,Stamper-Kurn00,Steinhauer02, Steinhauer03}, the structure factor of
molecular condensates \cite{Abo-Shaeer05} and to study signatures of vortices \cite{Blakie01,
Muniz06}. Furthermore, the technique is discussed for measuring the dynamic structure factor in the
presence of long-range particle correlations, such as superfluid pairing in a Fermi gas or fermionic
condensation \cite{Buechler04,Bruun06,Challis07}.

Because Bragg diffraction coherently couples two momentum states, it is frequently used as a
beamsplitter for matter waves in interferometric experiments \cite{Kasevich91}. A BEC seems the
optimal candidate for such an interferometer, because the macroscopically populated wavefunction
yields a high interferometric contrast. Consequently, a number of interferometric experiments have
been done using Bragg diffraction of a BEC, e.g. \cite{Simsarian00,Garcia06}. However, in the case
of a BEC, the two momentum states resulting from the Bragg diffraction interact with each other via
$s$-wave collisions \cite{Chikkatur00}. This interaction is prominently observed in time-of-flight
(TOF) absorption images as a halo forming between the diffraction peaks. The typical lifetime of
coherent superposition states due to interatomic interactions therefore is limited to only a few
tens of microseconds \cite{Inouye99}. Such experiments are thus performed during TOF or in very weak
traps.

Superposition states in a Fermi gas, however, can live for a very long time as has been observed for
potassium atoms in an optical lattice \cite{Roati04}. Since fermions do not interfere, the matter
wave contrast is necessarily limited to single-particle interference, but the long coherence
lifetime nevertheless makes fermions excellent candidates for interferometric experiments.

In this paper we report on Bragg diffraction of ultracold fermions from a light grating realized by
two slightly detuned counterpropagating laser beams. The diffraction was analyzed by studying Rabi
oscillations between the two coupled momentum states for low intensities of the Bragg lasers.
Increasing the intensity of the Bragg light eventually leads to Kapitza-Dirac scattering. We also
studied the transition between Bragg and Kapitza-Dirac scattering. Here, the system is no longer
well described as a two-level system and a multilevel ansatz has to be used. Experimental results of
scattering in the Kapitza-Dirac regime and their theoretical description are presented.

Furthermore, we used Bragg spectroscopy to map the momentum distribution of the $^6$Li cloud. As
there is no interaction between identical fermions in the ultracold regime, the momentum
distribution directly reflects the thermal distribution given by the Fermi-Dirac statistics. With a
temperature of $T/T_\mathrm F=0.6$ the onset of degeneracy can be found by comparing
Maxwell-Boltzmann and Fermi-Dirac fits to the measured momentum distribution \cite{DeMarco98}.

The diffracted momentum states oscillate in the trap for several seconds nearly unperturbed, which
is only possible because of suppression of $s$-wave collisions due to the Pauli principle. Higher
partial waves are frozen out at the ultralow temperatures realized in the experiment. The
oscillation is eventually destroyed by dephasing due to the anharmonicity of the trap. Furthermore,
the lifetime of oscillating atoms is dramatically reduced in the presence of $^{87}$Rb atoms that
collide with the $^6$Li atoms.

We also report on Ramsey interferometry with ultracold fermions, based on Bragg diffraction of
trapped $^6$Li atoms. A complete Ramsey spectrum can be taken in a single experimental cycle and is
recorded by absorption imaging after TOF. We observe Ramsey fringes for holding times of more than
30\,$\mu$s. For longer times, the fringes cannot be resolved by the imaging system.

The limitation can be overcome by echo techniques similar as in NMR \cite{Hahn50}. This method has
already been applied in atomic beam experiments \cite{Kasevich91} as well as in interferometric
experiments with BECs \cite{Garcia06}. In our experiment a diffraction echo is generated by a
two-photon Bragg pulse located halfway in time between the Ramsey pulses. Hence, we realize a
motional wavepacket echo by optical means in a similar way as it is done in a Ramsey-Bord\'e
interferometer \cite{Riehle91}. Our paper extends previous work presented in \cite{Marzok08}.

\section{Experimental procedure}
\label{SecExperiment}

To achieve simultaneous quantum degeneracy of $^6$Li and $^{87}$Rb, we use a procedure detailed in
previous papers \cite{Marzok08, Silber05}. $^6$Li atoms provided by a Zeeman slower and $^{87}$Rb
atoms ejected from dispensers \cite{Fortagh98} are simultaneously collected by a magneto-optical
trap. They are subsequently transferred via several intermediate magnetic traps into a
Ioffe-Pritchard type trap, where they are stored in their respective hyperfine states
$|F,m_F\rangle=|3/2,3/2\rangle$ and $|2,2\rangle$. This (compressed) trap is characterized by the
secular frequencies $(\omega_x,\omega_y,\omega_z)/2\pi=(762,762,190)\,\textrm{Hz}$ for $^6$Li and
the magnetic field offset $3.5\,$G. For $^{87}$Rb the trap frequencies are $\sqrt{87/6}$ times
lower. The $^{87}$Rb cloud is selectively cooled by microwave-induced forced evaporation and serves
as a cooling agent for the $^6$Li cloud, which adjusts its temperature to the $^{87}$Rb cloud
through interspecies thermalization. However, due to the small interspecies scattering length,
$a=-20\,a_B$ \cite{Silber05, Li08}, the thermalization is slow. Therefore, at low temperatures, when
the size of the evaporated $^{87}$Rb cloud becomes small or in shallow traps, where the different
gravitational sag separates the two species, the clouds thermally decouple. For the experiments
described below, we typically reach temperatures of below $1\,\mu\textrm{K}$ with $2.5\cdot10^6$
$^{87}$Rb atoms and $1.5\,\mu$K with $2\cdot10^5$ $^6$Li atoms. The respective densities are
$2\cdot10^{13}\,\textrm{cm}^{-3}$ for $^{87}$Rb and $4\cdot10^{12}\,\textrm{cm}^{-3}$ for $^6$Li,
which corresponds to the critical temperature $T_c=0.7\,\mu$K and the Fermi-temperature $T_\mathrm
F=2.4\,\mu$K.

    \begin{figure}[ht]
        \centerline{\includegraphics[width=12cm]{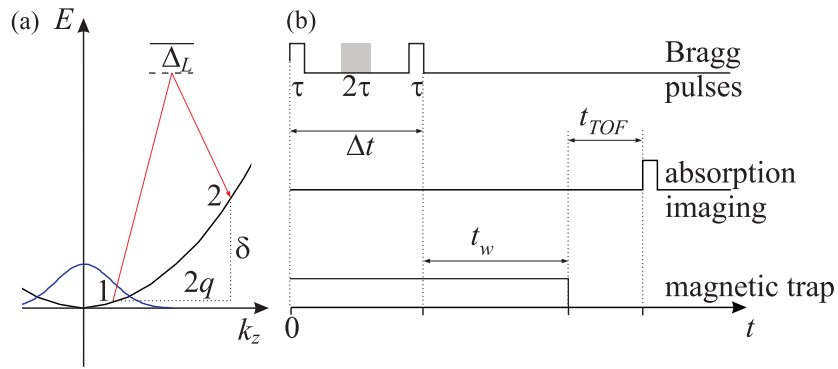}}
        \caption{Schematics of a Bragg diffraction experiment. (a) Dispersion parabola of (quasi-)
                free particles. Shown is the initial Gaussian momentum distribution of the atoms
                and the Bragg transition between the initial state $1$ and the final state $2$
                with a momentum shift of $2\hbar q$. (b) Typical pulse sequence of a Bragg diffraction
                experiment. For single-pulse experiments only the first Bragg pulse is applied. For a
                Ramsey experiment two $\pi/2$ pulses are used. For a diffraction echo experiment, a
                third pulse (shaded area) is centered between the Ramsey pulses.}
        \label{Fig01}
    \end{figure}

For most experiments (unless stated otherwise) we use a decompressed magnetic trap with trap
frequencies $(\omega_x,\omega_y,\omega_z)/2\pi =(236,180,141)\,\textrm{Hz}$. Decompression lowers
the temperatures of the clouds, but increases their gravitational sags. $^6$Li is much lighter than
$^{87}$Rb, so that the two clouds separate in space, and the $^6$Li cloud thermally decouples from
the $^{87}$Rb cloud. Since $^6$Li cannot thermalize by itself, the adiabatic cooling upon
decompression becomes anisotropic. The temperature measured along the $z$-axis, along which the
Bragg diffraction is performed, is then $T_z\simeq0.9\,\mu$K.

\subsection{Bragg diffraction}

The $^6$Li atoms are Bragg-diffracted by means of two counterpropagating laser beams aligned along
the $z$-axis of the Ioffe-Pritchard trap. The laser beams are frequency-shifted by acousto-optic
modulators (AOM). One AOM is driven by a stable quartz oscillator at 100\,MHz. The other AOM is
driven by a voltage-controlled oscillator, which is phase-locked to the quartz oscillator by means
of an electronic feedback loop. The frequencies of the two laser beams are set to differ by an
amount $\delta=\omega_2-\omega_1=2\hbar q^2/m=2\pi\cdot295\,$kHz, with $q=2\pi/\lambda$ and
$\lambda$ is the resonant wavelength of the $D_2$ line of $^6$Li. The Bragg beams have intensities
of $I_1=I_2 =13.7\,...\,132\,\textrm{mW/cm}^2$. The laser frequencies are tuned
$\Delta_L/2\pi\simeq1\,$GHz red to the $D_2$ line, so that the two-photon Rabi frequency reaches
values between $\Omega_\mathrm R=3\pi c^2\Gamma I/\hbar\omega^3\Delta_L\simeq2\pi\cdot
47\,...\,450\,$kHz. The pulse duration is chosen to generate specific pulse areas
$\Phi=\Omega_\mathrm R\tau/2\pi$. In most experiments, before applying the Bragg pulse, the
$^{87}$Rb atoms are removed from the trap by means of a resonant light pulse. After the Bragg pulse
we wait for a time $t_w$ before switching off the trapping field and recording absorption images
revealing the momentum distribution of the $^{87}$Rb and $^6$Li clouds. In some (Ramsey)
experiments, we apply a second Bragg pulse separated from the first one by a time interval $\Delta
t$, or even a third one at half time between the two others (see \fref{Fig01}).

    \begin{figure}[ht]
        \centerline{\includegraphics[width=9cm]{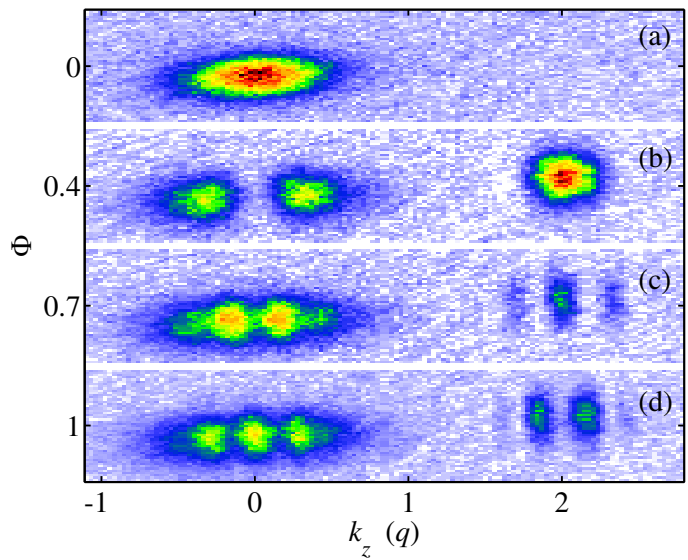}}
        \caption{TOF absorption images taken after a Bragg scattering pulse followed by a 4\,ms
                waiting time (half a trap period for re\-phasing) and a 2\,ms ballistic expansion.
                From (a) to (d) the Bragg pulse duration is varied such that the pulse area is
                $\Phi=\Omega_\mathrm R\tau/2\pi=0,\,0.4,\,0.7,\,1$.}
        \label{Fig02}
    \end{figure}

\subsection{Imaging}

\Fref{Fig02}(a-d) show typical $^6$Li absorption images taken after application of a Bragg pulse.
For low intensity of the Bragg beams, only a narrow slice is cut out of the fermionic momentum
distribution. The position of the slice along the $z$-axis depends on the detuning of the Bragg
lasers from the two-photon recoil shift, $\Delta=2\hbar q^2/m-\delta$, the width is due to power
broadening by the Rabi frequency $\Omega_\mathrm R$.

\section{Theoretical model}
\label{SecTheory}

\subsection{Free particles}

We describe our experimental observations with the following model \cite{Blakie00}. We assume that
during a Bragg pulse only two discrete momentum states $j=0,1$ of an atom are coupled and that the
impact of the trapping potential can be neglected. This is justified for sequences much shorter than
the trap oscillation period, i.e.\,$\tau,\Delta t\ll2\pi/\omega_z$.  The probability amplitudes of
the two states are denoted by $a_{j,k_z}$. They correspond to atoms with initial momentum $\hbar
k_z$ that are coupled to states with momentum $\hbar k_z'=\hbar(k_z+2q)$. The time evolution of the
amplitudes under the action of the Bragg light is given by the solution of the Schr\"odinger
equation
\begin{equation}\label{Eq01}
   \left(
   \begin{array}[c]{c}
       a_{0,k_z}(t) \\ a_{1,k_z}(t)
   \end{array}\right)
   = e^{-iH_{\tau}t/\hbar}\left(
   \begin{array}[c]{c}
       a_{0,k_z}(0) \\ a_{1,k_z}(0)
   \end{array}\right)\,,
\end{equation}
with the Hamiltonian
\begin{equation}\label{Eq02}
   H_{\tau} =\left(
   \begin{array}[c]{cc}
       \frac{\hbar}{2m}k_z^2 & \frac{1}{2}\Omega_\mathrm R \\
       \frac{1}{2}\Omega_\mathrm R & \frac{\hbar}{2m}k_z'^2-\delta
   \end{array}\right)\,.
\end{equation}
When the Bragg light is turned off, the Hamiltonian simplifies to
\begin{equation}\label{Eq03}
   H_{\Delta t} =\left(
   \begin{array}[c]{cc}
       \frac{\hbar}{2m}k_z^2 & 0 \\
       0 & \frac{\hbar}{2m}k_z'^2-\delta
   \end{array}\right)\,.
\end{equation}
By concatenating time evolutions described by $e^{-iH_{\tau}t/\hbar}$ and $e^{-iH_{\Delta
t}t/\hbar}$ the phase evolution of individual atoms in a superposition of momentum states can be
calculated for arbitrary sequences of pulses separated by times of free evolution, e.g.\,Ramsey type
sequences.

Initially the atoms are inhomogeneously distributed in momentum space. This distribution is governed
by Fermi-Dirac statistics. In the temperature range of our experiments ($T/T_\mathrm F\geq0.6$)
however, the distribution does not deviate much from Maxwell-Boltzmann statistics, such that the
momentum distribution can be assumed as
\begin{equation}\label{Eq04}
   \phi(k_z)=\hbar/\left(2\pi mk_\mathrm BT\right)^{1/2}e^{-\hbar^2k_z^2/2mk_\mathrm BT}\,.
\end{equation}
To obtain the momentum distribution of the atoms after an applied pulse sequence, we calculate the
evolution of the amplitudes $a_{j,k_z}$ for a variety of initial momenta and weight the final
populations of the momentum states with the distribution function $\phi(k_z)$. The number of atoms
in the zeroth and first order Bragg-diffracted modes is then
\begin{equation}\label{Eq05}
   N_j(t)=\int\phi(k_z)\left|a_{j,k_z}(t)\right|^2dk_z\,.
\end{equation}
The procedure neglects atomic interactions, which certainly is a good assumption for an ultracold
Fermi gas \cite{DeMarco99}.

\subsection{Trapped particles}

If trapped atoms are considered, the problem arises that the momentum eigenstates are simultaneously
coupled by two interactions, the moving optical lattice and the harmonic trap. However, the
situation is simplified, if a separation of scales is possible. In general, the duration of a pulse
is very short, $\tau\ll2\pi/\omega_z$. In contrast, the duration of a waiting time period $\Delta t$
can be such that it is no more negligible compared to a trap oscillation period. For these time
intervals, the action of the trapping potential must explicitly be taken into account.

Because the trap couples the atomic momenta and positions, the initial spatial distribution of the
atoms must now be considered. For simplicity we describe it as a thermal Gaussian function, similar
as done in \Eref{Eq04} for the momentum distribution,
\begin{equation}\label{Eq06}
   \psi(z) = \left(m\omega_z^2/2\pi k_\mathrm BT\right)^{1/2}e^{-m\omega_z^2z^2/2k_\mathrm BT}\,.
\end{equation}
Starting from initial positions $z$ and momenta $k_z$, a diffraction pulse transfers the recoil $2q$
to a part of the atoms. If the pulse length is short enough, the atomic position is not altered. The
atoms then follow the classical trajectories
\begin{eqnarray}\label{Eq07}
    \tilde k_z(t) & = & k_z\cos\omega_zt-\frac{\hbar\omega_z}{m}z\sin\omega_zt\,,\\
    \tilde k_z'(t) & = & (k_z+2q)\cos\omega_zt-\frac{\hbar\omega_z}{m}z\sin\omega_zt\,,\nonumber
\end{eqnarray}
where the first expression holds for undiffracted atoms and the second for diffracted atoms. These
values are used instead of $k_z$ and $k_z'$ respectively, in the Hamiltonians\,\eref{Eq02} and
\eref{Eq03}. As the Bragg pulses are short compared to the oscillation period,
$\tau\ll2\pi/\omega_z$, the effect of the trapping potential can be neglected for the description of
the Bragg diffraction. This means that the Hamiltonian $H_{\tau}$ depends on $\tilde k_z(t)$, but
can be treated as time-independent during the short time intervals $\tau$. In contrast, the free
propagation Hamiltonian $H_{\Delta t}$ gets time-dependent if $\Delta t$ is long. In this case, the
phase evolution of the atoms in both coupled momentum states can be described by writing the
time-evolution operator as
\begin{equation}\label{Eq09}
   e^{-iH_{\Delta t}t/\hbar}
    = \left(\begin{array}[c]{cc}
        \textrm{exp}\left(-i\int_0^t dt\frac{\hbar}{2m}\tilde k_z^2(t)\right) & 0 \\
        0 & \textrm{exp}\left(-i\int_0^t dt\left[\frac{\hbar}{2m}\tilde k_z'^2(t)-\delta\right]\right)
    \end{array}\right)\,.
\end{equation}

Since the amplitudes $a_{j,z,k_z}$ now also depend on the initial atomic positions, the final
populations of the momentum states must additionally be weighted by the initial spatial
distribution. Therefore, \Eref{Eq05} for the expectation values of the diffracted and non-diffracted
atom numbers now reads
\begin{equation}\label{Eq08}
   N_j(t)=\int\!\!\!\int\phi(k_z)\psi(z)\left|a_{j,z,k_z}(t)\right|^2dk_zdz\,.
\end{equation}

\section{Single pulse experiments}

\subsection{Bragg diffraction}

We first study diffraction with a single Bragg pulse. Having applied a pulse of a given length, we
observe an axial modulation of the momentum distribution. This is due to the fact that different
velocity classes of the atomic clouds have different Doppler shifts with respect to the Bragg lasers
and therefore are subject to different Rabi flopping frequencies. The initial momentum distribution
thus leads to an inhomogeneous population of the two momentum states after the pulse.
    \begin{figure}[hbtp]
        \centerline{\includegraphics[width=11cm]{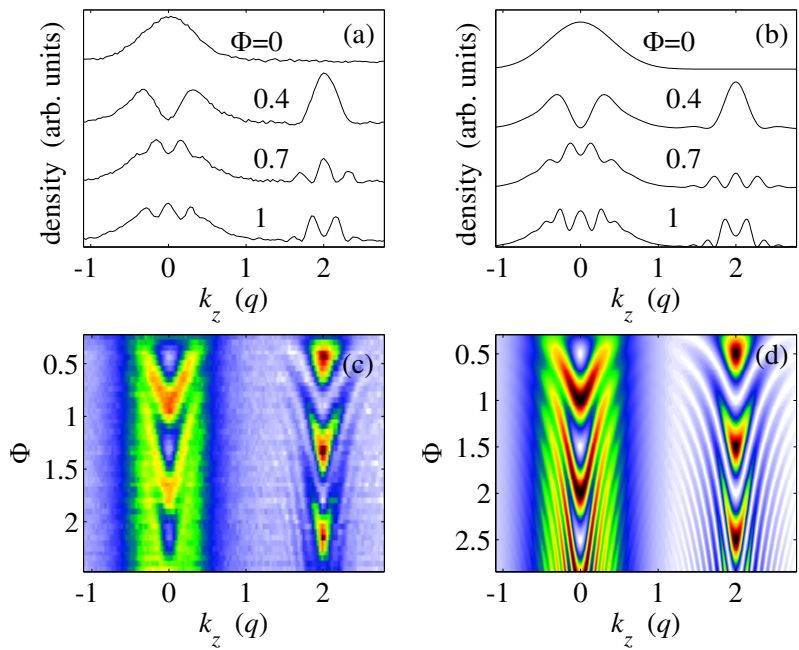}}
        \caption{(a) Integration of the TOF absorption images shown in \fref{Fig02} along the
                radial direction (perpendicular to $\mathbf{q}$). (b) Simulation of the axial momentum
                distribution after a Bragg pulse with same pulse area as in (a). c) False (color map
                of measured momentum distributions with pulse areas ranging between $\Phi=\Omega_\mathrm
                R\tau/2\pi=0\,...\,2.8$. The radially integrated absorption images appear as rows.
                (d) Calculation of the false color map with the experimental
                parameters used for the measurement (c).}
        \label{Fig03}
    \end{figure}

For small pulse areas $\Phi$ only momenta around $\hbar k_z=0$ are diffracted. With increasing
$\Phi$ nonresonant velocity classes are also coupled by the Bragg beams as can be seen in
\fref{Fig03}(a) and (c). The interpretation is supported by a theoretical simulation shown in
\fref{Fig03}(b) and (d), which has been done by computing the evolution of the momentum distribution
according to Equations \,\eref{Eq02} and \eref{Eq04}. Note, that the width of the initial momentum
distribution exceeds the power and Fourier broadening in this experiment.

\begin{figure}[ht]
        \centerline{\includegraphics[width=11cm]{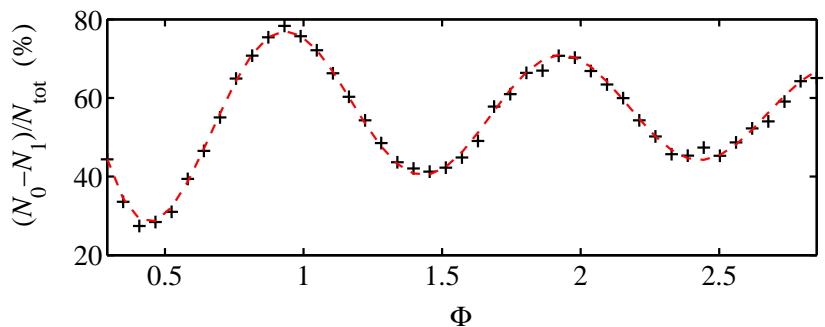}}
        \caption{Difference between the measured number of atoms in the zero and first-order
                Bragg-diffracted clouds for variable duration of the Bragg pulse. The (black) dots
                are the experimental data, the (red) dashed line is a theoretical simulation according
                to \Eref{Eq05}.}
        \label{Fig04}
    \end{figure}

The total amount of undiffracted and diffracted atoms is obtained by integration of the momentum
distributions in the zeroth and first-order Bragg-diffracted clouds, respectively. The measured
difference between the number of atoms in the first and zeroth order normalized to the total atom
number is shown in \fref{Fig04} (crosses) as a function of pulse duration. The observed damped Rabi
oscillations are reproduced by theory doing the same integration for the simulated momentum
distribution according to \Eref{Eq05}, as can be seen in \fref{Fig04} as a (red) dashed line. Note
that the damping is solely due to the initial inhomogeneous momentum distribution, decoherence was
not included in the simulations.

\subsection{Kapitza-Dirac scattering}

For larger Rabi frequencies,
\begin{equation}\label{Eq10}
    \Omega_\mathrm R\gg \frac{2\hbar q\sigma_{k_z}}{m}\,,
\end{equation}
with $\sigma_{k_z}=\sqrt{mk_\mathrm BT/\hbar^2}$ being the width of the momentum distribution, the
Doppler broadening is dominated by power broadening, meaning that all atoms can be diffracted into
the first order simultaneously.

At some point, however, the Rabi frequency becomes comparable to the energy difference between
adjacent momentum states, and Kapitza-Dirac scattering sets in. According to \cite{Blakie00} the
scattering will stay two-state like as long as the Rabi frequency fulfills the condition
\begin{equation}\label{Eq11}
    \Omega_\mathrm R\ll\frac{\hbar}{m}\left(4q^2-2q\sigma_{k_z}\right)\,.
\end{equation}

In the Kapitza-Dirac regime, the large energy uncertainty, connected with the fast coupling rate,
allows several momentum states to be coupled simultaneously. The resulting dynamics can be described
by a simple extension of the model detailed in \Sref{SecTheory}. In particular the
Hamiltonian\,\eref{Eq02} is replaced by
\begin{equation}\label{Eq12}
    H_{\tau}=\left(\begin{array}[c]{ccccc}
        \ddots & \ddots &  &  & \\
        \ddots & \frac{\hbar(k_z-2q)^2}{2m}+\delta & \frac{\Omega_\mathrm R}{2} &  & \\
        & \frac{\Omega_\mathrm R}{2} & \frac{\hbar k_z^2}{2m} & \frac{\Omega_\mathrm R}{2} & \\
        &  & \frac{\Omega_\mathrm R}{2} & \frac{\hbar(k_z+2q)^2}{2m}-\delta & \ddots\\
        &  &  & \ddots & \ddots
    \end{array}\right)\,.
\end{equation}
The crossover from the Bragg-diffraction regime to Kapitza-Dirac scattering is a smooth transition.
As we will see below, for intermediate Rabi frequencies (here $1\,\textrm{MHz}>\Omega_\mathrm
R/2\pi>100\,$kHz) the neighboring diffraction states, corresponding to momentum shifts of $4\hbar q$
and $-2\hbar q$ (second and minus first order), are scarcely populated. For increasingly higher Rabi
frequencies, $\Omega_\mathrm R/2\pi>1\,$MHz, the scattering populates more and more diffraction
orders.

\begin{figure}[ht]
        \centerline{\includegraphics[width=10cm]{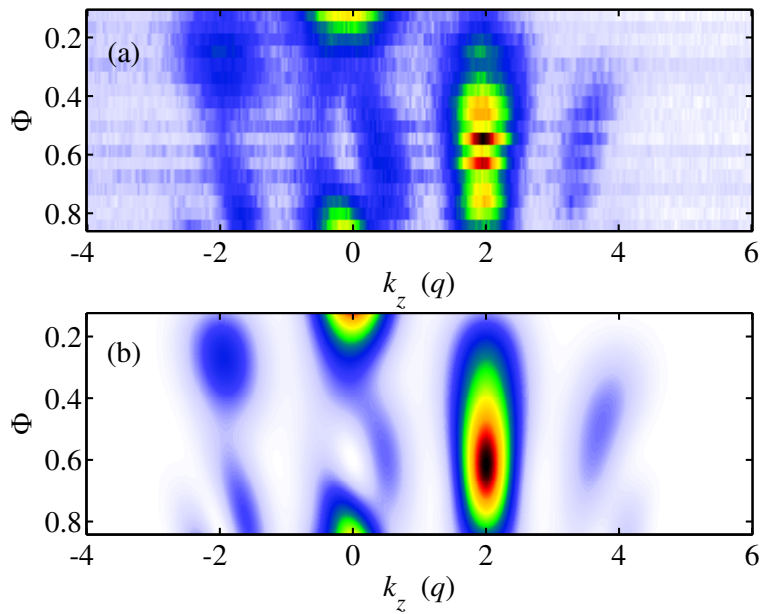}}
        \caption{(a) False color map  of measured momentum distributions similar to
                \fref{Fig03}. The Rabi frequency was chosen such that the system is approaching
                the Kapitza-Dirac regime. While the main diffraction is between the momentum states
                of $0\hbar q$ and $2\hbar q$ (zeroth and first order diffraction states), diffraction
                into the $-2\hbar q$ and $4\hbar q$ states also occur. The pulse area was changed by
                applying different pulse lengths. (b) Theoretical simulation of the momentum
                distribution in (a) using \Eref{Eq12}. The Rabi frequency chosen for the
                simulation was $\Omega_\mathrm R=2\pi\cdot420$\,kHz.}
        \label{Fig05}
    \end{figure}

We performed Bragg-diffraction at the transition between the Bragg and Kapitza-Dirac regime by
increasing the Bragg laser intensity beyond the value used in the experiments discussed in the
previous sections. \Fref{Fig05}(a) shows the atomic distribution as a function of $\Phi$ after 2 ms
TOF for such an experiment. A simulation using \Eref{Eq12} is shown in \fref{Fig05}(b). This
simulation reproduces the experimental results very well and reveals a Rabi frequency of
$\Omega_\mathrm R=2\pi\cdot420$\,kHz.

In some cases it is desirable to diffract all atoms simultaneously into the first order. This is
only feasible by combining the Bragg regime \eref{Eq11}, $\Omega_\mathrm
R\ll\frac{\hbar}{m}\left(4q^2-2q\sigma_{k_z}\right)$ with large power broadening \eref{Eq10},
$\Omega_\mathrm R\gg 2\hbar q\sigma_{k_z}/m$. The resulting condition,
\begin{equation}\label{Eq13}
    k_\mathrm BT\ll\frac{\hbar^2q^2}{m}\,,
\end{equation}
is satisfied for temperatures $T\ll7\,\mu$K. The temperature reached in this experiment are within
this regime.

\subsection{Bragg spectroscopy}

For a given frequency detuning $\delta$ the Bragg lasers address a specific momentum class $\hbar
k_z$ of the atomic cloud \cite{Stenger99},
\begin{equation}\label{Eq14}
    \delta=\frac{2\hbar q^2}{m}+\frac{2\hbar qk_z}{m}\,.
\end{equation}
This implies a linear dependence between detuning and addressed momentum class. It is thus possible
to probe the momentum distribution by measuring the number of Bragg-diffracted atoms as a function
of the detuning.

    \begin{figure}[ht]
        \centerline{\includegraphics[width=9cm]{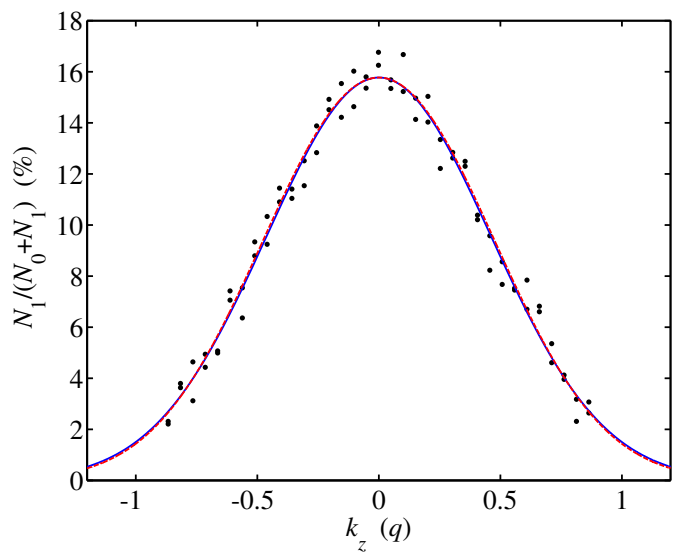}}
        \caption{Momentum distribution of the trapped $^6$Li cloud measured by Bragg spectroscopy.
                The blue (solid) line is a Gaussian fit to the data, revealing a temperature of
                $1.5\,\mu$K. The red (dashed) line is a simulated Fermi-Dirac momentum distribution
                for $2\cdot10^5$ atoms at $1.4\,\mu$K. These two curves are nearly indistinguishable
                in the plot.}
        \label{Fig06}
    \end{figure}

\Fref{Fig06} shows  a spectrum resulting from such a series of measurements, performed in the
compressed trap. In this experiment, the pulse duration was chosen such that the pulse area was
$\Phi\simeq0.5$. The frequency detuning of the two Bragg lasers was changed between 40\,kHz and
550\,kHz in 15\,kHz steps, which is small compared to the Rabi frequency $\Omega_\mathrm
R\simeq2\pi\cdot58$\,kHz. In \fref{Fig06} the amount of scattered atoms is plotted versus the
resonant atomic momentum derived from \Eref{Eq14}. The data show a distinctive Gaussian shape,
corresponding to the temperature $T=1.50\,\mu\textrm{K}\pm0.15\,\mu$K (blue, solid curve), in good
agreement with TOF measurements.

At low temperatures $T\ll T_\mathrm F$, the momentum distribution is expected to deviate from a
Gaussian profile and to adopt the shape of an inverted parabola. However, in our experiments the
temperatures were not far below the Fermi temperature of $T_\mathrm F\simeq2.4\,\mu$K. In this
regime, the distributions, integrated in the two dimensions perpendicular to the $z$-axis, have
quite similar shapes. A fit based on a Fermi-Dirac distribution for $1.8\cdot10^5$ atoms is also
shown in \fref{Fig06} as red (dashed) curve. This fit yields a temperature of
$T=1.39\,\mu\textrm{K}\pm0.18\,\mu$K.

The two curves, for the Gaussian and the Fermi-Dirac distribution are nearly congruent. Although the
shape of the momentum distribution thus provides no clear signal for degeneracy in this regime, the
result suggests that the temperatures due to Gaussian fits are overestimated by 5..10\,\%.

Bragg spectroscopy yields the same information as standard TOF absorption imaging. However, this
technique reveals the momentum distribution inside the trap. Adiabatic cooling and magnetic field
distortions during trap switch-off do not affect the signal, which is particularly important for
light and fast atoms such as lithium.

\subsection{Trap dynamics}
\label{Sec4.4}

\begin{figure}[ht]
        \centerline{\includegraphics[width=12cm]{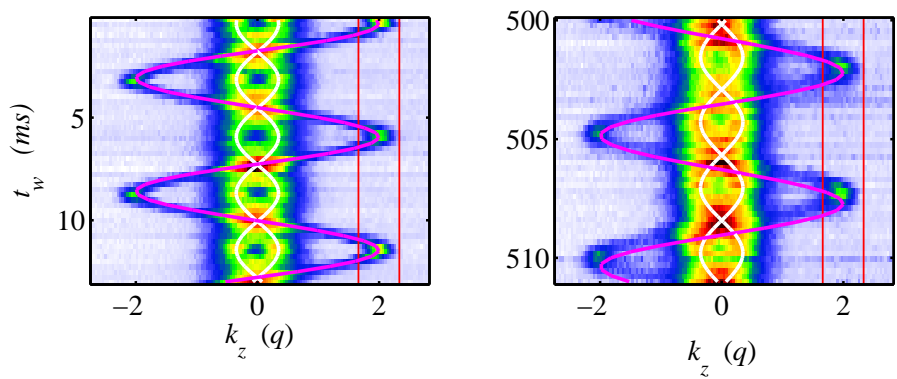}}
        \caption{Momentum distribution after a $\pi$-pulse and a variable waiting time $t_w$ in
                the trap. Diffracted and undiffracted atoms oscillate in a potential given in good
                approximation by \Eref{Eq10b}. The equation of motion for this oscillation
                has been computed and is shown as (white and magenta) curves. The oscillation period
                for the two amplitudes differ, as can be seen for large waiting times
                $t_w\gg2\pi/\omega_z$. The (red) lines indicate the area in which the atom numbers
                were counted for analysis in \fref{Fig08}.}
        \label{Fig07}
    \end{figure}

By Bragg diffraction the atoms acquire additional momentum and begin to oscillate inside the trap.
This oscillation has been analyzed in the following experiment. A fraction of the atoms, stored in
the compressed trap, is Bragg-scattered by a single $\pi$-pulse. The atoms are then held inside the
trap for variable holding times $t_w$ before being imaged (see \fref{Fig07}).

Several observations are made: If no $^{87}$Rb is in the trap, the $^6$Li atoms oscillate for times
on the order of seconds; however, the distribution of the diffracted atoms smears out. Also, the
oscillation period of the diffracted atoms differs from the period of undiffracted atoms. The
momentum distribution of these undiffracted atoms is depleted around $k_z=0$, because atoms in low
momentum states are Bragg-scattered preferentially. Atoms in the zeroth order oscillate with an
average amplitude given by the width of the thermal distribution of the original cloud, as can be
seen in \fref{Fig07}.

The broadening of the momentum distribution as well as the different oscillation periods for the two
diffraction orders are due to anharmonicities  in the magnetic trap. The magnetic potential of the
trap is a combination of the quadrupole field generated by a pair of coils in anti-Helmholtz
configuration and the field generated by four wires parallel to the rotational axes of the coils.
The numerically calculated field geometry is in very good approximation given by
\cite{Silberthesis06}
\begin{equation}\label{Eq10b}
    B_z(z)=B_0-\alpha(z+a)+\beta a^2e^{z/a}\,,
\end{equation}
with the gradient $\alpha$, the curvature $\beta$ and $a=\alpha/\beta$. With the resulting
potential, the oscillation of the atoms in the cloud can be numerically simulated and compared to
the anharmonic motion observed in the experiment. From this we derive values of
$\alpha=135\,\textrm{G}/\textrm{cm}\pm25\,\textrm{G}/$cm for the gradient and
$\beta=1.402\cdot10^3\textrm{G}/\textrm{cm}^2\pm 4\,\textrm{G}/\textrm{cm}^2$ for the curvature,
leading to the trap frequency $\omega_z=\sqrt{\mu_\mathrm B\beta_z/m}=2\pi\cdot181.6\,$Hz. The
numerical calculation of the magnetic field configuration confirms the measured gradient
\cite{Silberthesis06}. Simple harmonic fits to the oscillation result in the same oscillation
frequency with an error better than $10^{-2}$. This implies that the effect of the anharmonicity on
the atomic motion is less than $1\%$.

    \begin{figure}[ht]
        \centerline{\includegraphics[width=10cm]{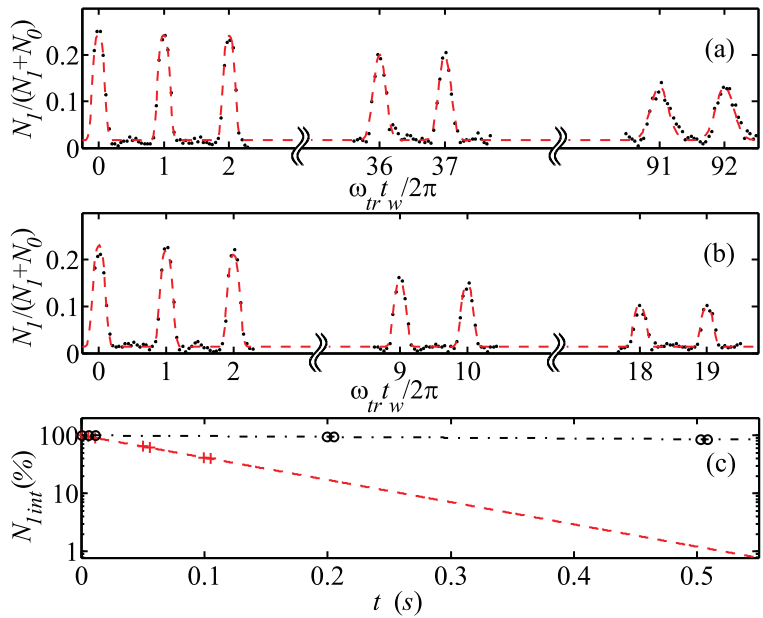}}
        \caption{Rephasing of the momentum distribution in the trap in the absence (a) and in the
                presence (b) of $^{87}$Rb. Plotted are the integrated numbers of Bragg-diffracted
                atoms found at a given time $t_w$ after an initial Bragg $\pi$-pulse in a restricted
                area of the momentum space. The fit is a combination of an error function with linearly
                increasing Gaussian width and an exponential damping of the peak height. This damping
                is best observed in (c), where the integrated number of atoms in each peak is plotted
                versus time in a logarithmic plot. Only small damping is observed in the absence of
                $^{87}$Rb (black, dash-dotted curve), while the presence of $^{87}$Rb induces a fast
                damping (red, dashed curve).}
        \label{Fig08}
    \end{figure}

\bigskip

We also analyzed the data by counting the number of atoms in a certain area of momentum space. This
area was chosen as a box completely encompassing the diffracted atomic cloud at their highest
momentum as indicated by the (red) lines in \fref{Fig07}. The result of this analysis is shown in
\fref{Fig08}. Here, the fraction of atoms in the evaluation area relative to the total number of
atoms is plotted as a function of time. The peaks are separated by a trap period and represent a
revival of the original momentum distribution, directly after the Bragg pulse. For the fit in
\fref{Fig08} (red dashed curve) we assume a Gaussian cloud shape oscillating in and out of the
detection volume. The convolution of the Gaussian and the rectangular shape of the detection volume
is described by a combination of two error functions with opposite slopes shifted with respect to
each other by the time it takes for the cloud to oscillate in and out of the detection volume. The
steepness of the slopes is given by the width of the cloud, and this feature is repeated every trap
period.

As one can see from \fref{Fig08}(a), the width $\sigma$ of the cloud's Gaussian momentum
distribution broadens with time. If we assume a linear broadening, the fit to the data yields
\begin{equation}
    \sigma(t)=\sigma_0(1+5\cdot10^{-6}\frac{t}{\mu\textrm{s}})\,
\end{equation}
where $\sigma_0$ is the initial width. This broadening is the main reason for the reduction in
amplitude and is due to the anharmonicity of the trap. The integrated number of atoms in each peak,
on the other hand, decreases much slower. This can be seen in \fref{Fig08}(c), where the integrated
number of atoms per peak is plotted against time. In the absence of $^{87}$Rb we find a damping time
of 3.2\,s corresponding to more than 500 oscillations \footnote{In the decompressed trap, we found
damping times without $^{87}$Rb atoms exceeding 13\,s, which is on the order of the
    magnetic trap lifetime. However, in those traps the presence of a $^{87}$Rb cloud has no impact on
    the $^6$Li because of gravitational
    sag.}.

The exponential decrease of the atom number is due to $p$-wave collisions between diffracted and
undiffracted atoms, since $s$-wave collisions are forbidden for a spin-polarized Fermi gas. The
damping time therefore corresponds to a collision rate of $\gamma_{\mathrm{coll\,6}p}=0.3\,$Hz. From
this the $p$-wave scattering cross section can be calculated \cite{Silber05,Marzok07}. Here we
assume a Gaussian density distribution for the undiffracted atoms and a harmonic oscillation of the
diffracted atoms. The relative scattering velocity $v_{\mathrm{sc}}=2\hbar q/m$ in this system
corresponds to a scattering energy of $5\,\mu\textrm{K}\cdot k_\mathrm B$. We derive a $p$-wave
scattering cross section of $\sigma_p=1.4\cdot10^{-13}\,\textrm{cm}^2$, which agrees with similar
measurements performed on $^{40}$K \cite{DeMarco99} and calculations made for $^6$Li \cite{Chevy05}.

In the presence of a cloud of $^{87}$Rb, we find a dramatic reduction of the damping time, as can be
seen in \fref{Fig08}(b) and (c). The $^6$Li atoms may collide with $^{87}$Rb atoms if these are not
initially removed from the trap. The collision time constant calculated from the heteronuclear
$s$-wave scattering length \cite{Silber05} is $\gamma_{\mathrm{coll\,6,87}}^{-1}=107\,$ms. The
exponential damping in the fit of \fref{Fig08}(b) and (c) resulted in a damping time of 113\,ms,
while the broadening is of the same order as in the case without $^{87}$Rb. There is excellent
agreement between the calculated collision rate and the observed damping.

\section{Atomic coherences}

\subsection{Ramsey interferometry}

The observed revivals of the density distributions after long times indicate that the $^6$Li atoms
are free from any kind of perturbation, which in turn suggests long decoherence times for quantum
mechanical superpositions. The existence and the stability of such states can be tested by
interferometric experiments.

\begin{figure}[ht]
        \centerline{\includegraphics[width=10cm]{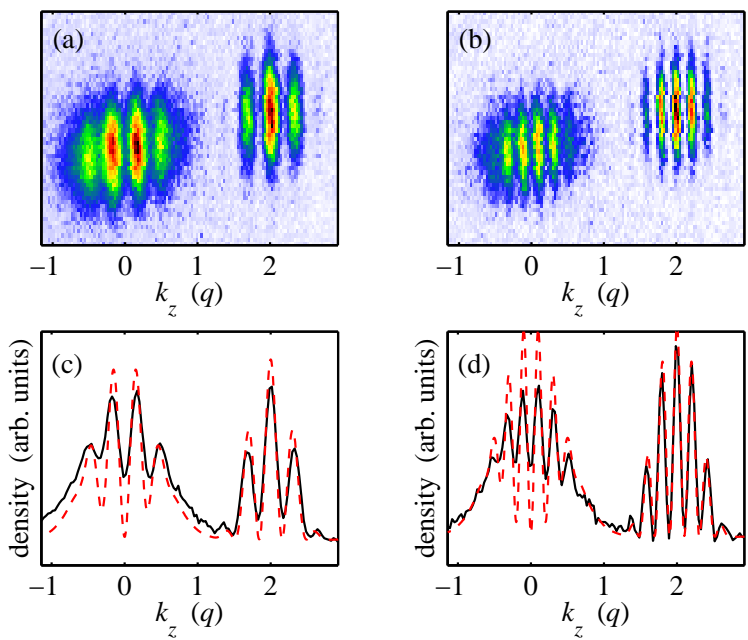}}
        \caption{(a,b) Absorption images taken after a Ramsey pulse sequence and 2\,ms ballistic
                expansion. (c,d) Integrations of the absorption images perpendicular to $k_z$. The
                (black) solid line is the experimental data, the (red) dashed line is a theoretical
                simulation using Equations \eref{Eq02}, \eref{Eq03} and \eref{Eq04}. The evolution
                time between the two Ramsey pulses is $\Delta t=6\,\mu$s in (a,c) and
                $\Delta t=12\,\mu$s in (b,d). The temperature used for the simulation is $T=1.2\,\mu$K
                and the Rabi frequency is $\Omega_\mathrm R=2\pi\cdot58\,$kHz.}
        \label{Fig09}
    \end{figure}

To this end, we apply a Ramsey sequence of two $\pi/2$ pulses, like in an atomic Mach-Zehnder (or
Ramsey-Bord\'e) type interferometer. The beamsplitters are Bragg pulses with a duration equivalent
to a pulse area of $\Phi=0.25$, and the interferometric paths are closed by the kinetic energy
released during time-of-flight. We proceed as follows. At time $t=0$ in the decompressed trap, a
first $\pi/2$ pulse of duration $\tau$ is applied. Then we wait for a variable amount of time
$\Delta t$ after which a second $\pi/2$ pulse is applied. The momentum distribution is recorded
after the delay of half a trap cycle and 2\,ms ballistic expansion. Typical time-of-flight
absorption images are shown in \fref{Fig09} (a) and (b).

The initial momentum distribution of the atoms gives rise to an inhomogeneous Doppler shift, which
detunes the atoms from Bragg resonance. This allows to observe all Ramsey fringes in a single shot.
We do observe Ramsey fringes for time separations of the Ramsey pulses on the order of up to
$30\,\mu$s. For longer times, the fringe spacing falls below the resolution limit of the imaging
system.

Using \Eref{Eq03} the phase difference for the two momentum states of an atom moving at velocity
$\hbar k_z/m$ accumulated after a time $\Delta t$ is given by
\begin{equation}
    \phi=\frac{2\hbar(q+k_z)q}{m}\Delta t-\delta\,.
\end{equation}
Adjacent fringes are produced by atoms whose velocities differ by an amount $\hbar\Delta k_z/m$ such
that $\Delta\phi=2\pi$, yielding the condition
\begin{equation}
    \frac{\hbar\Delta k_z}{m}=\frac{\pi}{q\Delta t}\,.
\end{equation}
After ballistic expansion time during a time $t_\mathrm{tof}$ the fringes are separated in space by
an amount $\pi t_\mathrm{tof}/q\Delta t$, which must exceed the $\Delta x=14\,\mu$m resolution of
our optical system. From this follows the condition for the maximum allowable Ramsey pulse
separation,
\begin{equation}
    \Delta t_\mathrm{max}=\frac{\pi t_\mathrm{tof}}{\Delta x\cdot q}=48\,\mu\textrm{s}\,.
\end{equation}
The realistic value is additionally subject to noise.

To a certain extent the contrast can be enhanced by longer TOFs. However this also dilutes the cloud
and reduces the quality of the absorption image. Lower temperatures may allow for longer TOFs, but
as the temperature of the cloud is below the Fermi temperature already, the width of the momentum
distribution cannot be reduced by much due to the Pauli exclusion principle.

The maximum detectable evolution time could be slightly improved by a better imaging system,
however, the general problem remains. It can be circumvented by an echo technique described in the
following section.

\subsection{Diffraction echo interferometry}

Interferometry of particle ensembles is subject to decoherence and dephasing. Decoherence destroys
quantum mechanical superpositions due to coupling to a reservoir. In contrast, dephasing results
from divergent phase space trajectories for different particles and may be reversed if the
trajectories can be inverted \cite{Andersen03}. This is possible with echo techniques well known in
NMR spectroscopy \cite{Hahn50}. They have been successfully applied in cold atom experiments
\cite{Kasevich91} and BEC interferometry \cite{Garcia06}. In an echo experiment the free evolution
time between two Ramsey pulses is split in two intervals by an additional $\pi$-pulse inverting the
phase evolution of atomic states. Diffraction echo interferometry is robust against dephasing.

    \begin{figure}[ht]
        \centerline{\includegraphics[width=9cm]{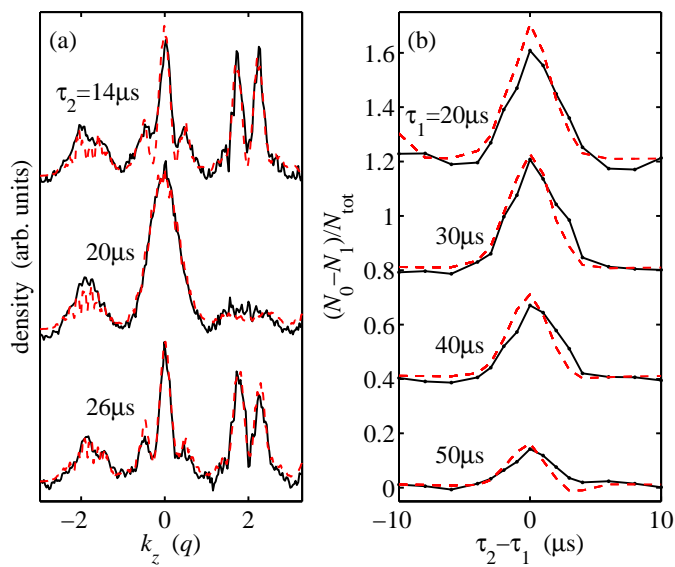}}
        \caption{(a) Integrated density of absorption images taken after a spin echo pulse sequence
                and 2\,ms TOF, for a waiting time $\tau_1 = 20\,\mu$s and a variable waiting time
                $\tau_2$. (b) Deviation between the total number of atoms in the zeroth and first
                exited state. The theoretical curves in red (dashed) are simulations based on
                \Eref{Eq09}, using a Rabi frequency of $\Omega_\mathrm R=2\pi\cdot220$\,kHz.}
        \label{Fig10}
    \end{figure}

In our case the signature for diffraction echo is the almost complete revival of the zeroth order
momentum state. Such an experiment is shown in \fref{Fig10}. It is similar to the Ramsey experiment
described above, but the waiting  time between the pulses is now divided into two waiting times
$\tau_1$ ($20\,\mu$s in \fref{Fig10}(a)) and $\tau_2$, separated by an additional $\pi$-pulse. After
completion of a pulse sequence, the atoms are held in the trap for half a trap period, then the trap
is turned off and the atoms are imaged after 2\,ms ballistic expansion time.

The Rabi frequency in this experiment is $\Omega_\mathrm R=2\pi\cdot220\,$kHz. This ensures Bragg
resonance for all atoms according to Equation \eref{Eq10}. However, for this Rabi frequency
Kapitza-Dirac scattering already occurs, as can be seen in \fref{Fig10}. For times
$\tau_2\neq\tau_1$ the zero-momentum revival is incomplete, it appears as a central peak (dip) in
the zeroth (first) order Bragg-diffracted portion. This peak (dip) broadens until for
$\tau_2=\tau_1$ the revival is most pronounced. This effect is also seen in \fref{Fig10}(b), where
the difference between atom numbers in the zeroth and first diffracted state is shown.

The revival is seen for times up to $\tau_{\mathrm{tot}}=\tau_1+\tau_2$=100\,$\mu$s. In a harmonic
trapping potential momentum states are no eigenstates, which leads to a mixing of position and
momentum coordinates for long evolution times. This process is responsible for the decay of the
diffraction echo rather than decoherence. The coherent theoretical model based on \Eref{Eq09} fully
describes the experimental observation, which leads us to claim $\tau_{\mathrm{tot}}$ as a lower
bound for the coherence time.

Setting the waiting times to a multiple of the oscillation period $\tau_1=\tau_2=2\pi n/\omega_z$
can compensate the effect of the trapping  potential as has been shown in \cite{Horikoshi07}. In our
experimental setup, revivals of the diffraction echo could not be observed. This might be due to the
large momentum spread of the initial cloud.

\section{Conclusion}

In conclusion, we presented Bragg scattering of fermions from a moving optical lattice. We were able
to show the coherence between the zeroth and first order diffracted momentum states. Our work
presents the first study of interferometry with fermionic atoms using separate interferometric arms
and a variable evolution time. The coherence is also observed with a diffraction echo experiment,
where a lower bound for the coherence time of $t_{\mathrm{tot}}>100\,\mu$s was found. A longer
observation was hampered by the fact that the studied momentum states are no eigenstates of the
harmonic trapping potential. The diffracted atoms, however, oscillate in the trapping potential for
very long times with a single atom damping time of about 3.2\,s. This leads us to the assumption of
a similarly long coherence time, which is consistent with the observation made by Roati \textit{et
al.} \cite{Roati04}. Thus fermionic $^6$Li is a promising candidate for interferometric experiments
with long coherence times using schemes similar to \cite{Garcia06}.

Moreover, we successfully applied Bragg spectroscopy to fermions, measuring the momentum
distribution of an ultracold degenerate Fermi gas. Although not deep in the degenerate regime
($T/T_\mathrm F=0.6$), the momentum distribution revealed signs of quantum degeneracy. For even
lower temperatures it is conceivable to observe degeneracy from the shape of the distribution
directly. In addition, Bragg spectroscopy allows for \textit{in-situ} probing of specific momentum
classes of trapped atoms.

Future work will be devoted to studying the $^{87}$Rb$+^6$Li mixture in a crossed beam dipole trap,
using Bragg scattering in the vicinity of one of the recently found heteronuclear Feshbach
resonances \cite{Deh08}. Signatures of the interaction energy of the ensemble should be visible in
the profile of a Bragg spectrum, similar to the frequency shift observed in Bragg spectroscopy on
BECs \cite{Stenger99}. Furthermore, Bragg spectroscopy permits to measure the atomic dispersion
relation, which is particularly interesting in case of many-body correlations. Bragg spectroscopy
should also be applicable to take excitation spectra of fermionic spin mixtures in the BEC-BCS
crossover regime which could be realized in our experiment as well.

\bigskip

\section*{Acknowledgements}
\addcontentsline{toc}{section}{Acknowledgements}

During the completion of this paper, we learned that Bragg spectroscopy on fermionic spin mixtures
has been done in Tokyo \cite{Inada08} and Melbourne \cite{Veeravally08}. This work has been
supported by the Deutsche Forschungsgemeinschaft (DFG).

\end{document}